\documentclass[lettersize,journal]{IEEEtran}
\usepackage{amsmath,amsfonts}
\usepackage{algorithm}
\usepackage{array}
\usepackage[caption=false,font=normalsize,labelfont=sf,textfont=sf]{subfig}
\usepackage{textcomp}
\usepackage{stfloats}
\usepackage{url}
\usepackage{verbatim}
\usepackage{graphicx}
\usepackage{cite}
\usepackage{multirow}
\usepackage{booktabs}
\usepackage{pifont}
\usepackage[pdftex,dvipsnames]{xcolor}

\usepackage[colorinlistoftodos,prependcaption,textsize=tiny]{todonotes}

\definecolor{mygray}{gray}{.9}

\usepackage{xargs}
\usepackage[colorinlistoftodos,prependcaption,textsize=tiny]{todonotes}
\newcommandx{\xy}[2][1=]{\todo[linecolor=red,backgroundcolor=red!25,bordercolor=red,#1]{#2}}
\newcommandx{\change}[2][1=]{\todo[linecolor=blue,backgroundcolor=blue!25,bordercolor=blue,#1]{#2}}
\newcommandx{\info}[2][1=]{\todo[linecolor=OliveGreen,backgroundcolor=OliveGreen!25,bordercolor=OliveGreen,#1]{#2}}
\newcommandx{\improvement}[2][1=]{\todo[linecolor=Plum,backgroundcolor=Plum!25,bordercolor=Plum,#1]{#2}}
\newcommandx{\thiswillnotshow}[2][1=]{\todo[disable,#1]{#2}}

\definecolor{mygray}{gray}{.9}
\newcommand{\stdv}[1]{\scalebox{.82}{~$\pm$~#1}}

\usepackage{graphicx}
\usepackage{multirow} 
\usepackage{tcolorbox}
\usepackage{amsmath}
\usepackage{amssymb}
\usepackage{multirow}
\usepackage{graphics}
\usepackage{threeparttable}
\usepackage{color}
\usepackage[normalem]{ulem}
\usepackage{float}
\usepackage{amsfonts}
\usepackage{bm}
\usepackage{bbm}
\usepackage{enumitem}
\usepackage{wrapfig}
\usepackage{subcaption}
\usepackage{algorithm}
\usepackage{array}
\usepackage{makecell}
\usepackage{color, colortbl}
\usepackage{pifont}
\usepackage{booktabs}
\usepackage{mathtools}
\usepackage{siunitx}
\usepackage{caption} 
\usepackage{subcaption} 
\usepackage[nopar]{lipsum}
\usepackage{listings}
\usepackage[export]{adjustbox}
\usepackage{makecell}
\usepackage{wrapfig}
\usepackage{mathtools}
\usepackage{siunitx}
\usepackage[nopar]{lipsum}
\usepackage{listings}

\usepackage{enumitem}
\usepackage{algpseudocode}
\usepackage[font=small,labelfont=bf]{caption}
\usepackage{array}
\usepackage{multirow}
\usepackage{booktabs}
\usepackage{algorithm}
\usepackage{subcaption}
\usepackage[normalem]{ulem}
\usepackage{xparse}
\usepackage{pifont}
\usepackage{bm}
\usepackage{threeparttable}
\usepackage{etoolbox}

\usepackage{listings}
\usepackage{mwe} %
\usepackage{makecell}
\usepackage{color, colortbl}
\usepackage{tabularx}
\usepackage{pifont}%
\usepackage[accsupp]{axessibility}
\usepackage{listings}
\definecolor{myy}{RGB}{126,95,0}
\definecolor{mygray}{gray}{.9}
\definecolor{Gray}{gray}{0.9}
\definecolor{bblue}{RGB}{30,80,120}
\definecolor{mygray1}{gray}{.7}
\definecolor{ggray}{RGB}{127,127,127}
\definecolor{defaultcolor}{gray}{.9}
\definecolor{dark-gray}{gray}{0.20}

\definecolor{mygreen}{HTML}{39b54a}

\newcolumntype{x}[1]{>{\centering\arraybackslash}p{#1pt}}
\newcolumntype{y}[1]{>{\raggedright\arraybackslash}p{#1pt}}
\newcolumntype{z}[1]{>{\raggedleft\arraybackslash}p{#1pt}}

\definecolor{cvprblue}{rgb}{0.21,0.49,0.74}
\usepackage{color}

\definecolor{micoblue}{HTML}{155AE5}
\definecolor{micopurple}{HTML}{B060E2}
\definecolor{micogreen}{HTML}{8FE260}
\definecolor{micoyellow}{HTML}{DEE260}
\definecolor{micored}{HTML}{E26060}

\usepackage{enumitem}
\usepackage{algpseudocode}
\usepackage[font=small,labelfont=bf]{caption}
\usepackage{array}
\usepackage{multirow}
\usepackage{booktabs}
\usepackage{subcaption}
\usepackage[normalem]{ulem}
\usepackage{xparse}
\usepackage{pifont}
\usepackage{bm}
\usepackage{threeparttable}
\usepackage{etoolbox}
\usepackage{listings}
\usepackage{mwe} %
\usepackage{makecell}
\usepackage{color, colortbl}
\usepackage{tabularx}
\usepackage{pifont}%
\usepackage[accsupp]{axessibility}

\usepackage[scaled=0.85]{DejaVuSansMono}

\definecolor{citecolor}{HTML}{0071bc}

\usepackage{graphicx}
\usepackage{amsmath}
\usepackage{amssymb}
\usepackage{pgfplots}
\usepackage{multirow}
\pgfplotsset{compat=1.16}
\usepgfplotslibrary{groupplots}
\usetikzlibrary{patterns}

\newlength\savewidth

\newlength\thinwidth
\definecolor{Gray}{gray}{0.92}
\definecolor{DarkGray}{gray}{0.5}
\newcolumntype{H}{>{\setbox0=\hbox\bgroup}c<{\egroup}@{}}
\definecolor{LightCyan}{rgb}{0.88,1,1}
\definecolor{altRowColor}{gray}{0.92}
\definecolor{highlightRowColor}{rgb}{0.9, 0.9, 1}

\definecolor{GrayNumber}{gray}{0.5}
\definecolor{GrayXMark}{gray}{0.7}
\definecolor{ER}{rgb}{1,0.27,0}
\definecolor{MI}{rgb}{0,0.3,0.8}
\definecolor{MW}{rgb}{.5,.0,.5}
\definecolor{SS}{rgb}{0,.5,0}
\definecolor{CM}{rgb}{0.992156862745098, 0.7686274509803922, 0.0941176470588235}
\definecolor{darkgreen}{RGB}{50,200,0}
\definecolor{lightblue}{HTML}{dfebf7}

\definecolor{firstbest}{RGB}{152,220,220} % 浅蓝色 (LightBlue)
\definecolor{secondbest}{RGB}{224,255,255} % 淡青色 (LightCyan)

\definecolor{ThermalDark}{rgb}{0.8823529411,0.63725490196,0.0156862745}
\definecolor{IMUDark}{rgb}{0.6235294117647059, 0.27058823529411763, 0.4627450980392157}
\definecolor{TimeDark}{HTML}{45DC61}  
\definecolor{GraphDark}{HTML}{00A6FC}
\definecolor{HyperDark}{HTML}{D1A44F}
\definecolor{TabularDark}{HTML}{FF7900}
\newcolumntype{i}{>{\columncolor{Image}}c}
\newcolumntype{v}{>{\columncolor{Video}}c}
\newcolumntype{d}{>{\columncolor{Depth}}c}
\newcolumntype{a}{>{\columncolor{Audio}}c}
\newcolumntype{I}{>{\columncolor{ImageLight}}c}
\newcolumntype{D}{>{\columncolor{DepthLight}}c}
\newcolumntype{A}{>{\columncolor{AudioLight}}c}
\newcolumntype{E}{>{\columncolor{highlightRowColor}}c}

\begin{document}

\title{\raisebox{-3pt}[0pt][0pt]{\includegraphics[width=0.06\textwidth]{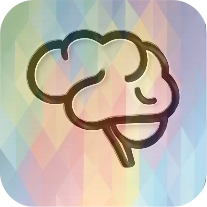}} BrainGPT: Unleashing the Potential of EEG Generalist Foundation Model by Autoregressive Pre-training}

\author{Tongtian Yue, Xuange Gao, Shuning Xue, Yepeng Tang, Longteng Guo, Jie Jiang, Jing Liu, \textit{Member, IEEE}

\thanks{
 This research is supported by Artificial Intelligence National Science and Technology Major Project under Grant No. 2023ZD0121200, National Natural Science Foundation of China under Grant No. 62437001, 62436001, Strategic Priority Research Program of the Chinese Academy of Sciences under Grant No. XDA0460305, and Key Research and Development Program of Jiangsu Province under Grant No. BE2023016-3. \textit{(Corresponding author: Jing Liu.)}}

\thanks{
Tongtian Yue, Xuange Gao, Shuning Xue and Yepeng Tang contributed equally to this work. Tongtian Yue, Xuange Gao, Shuning Xue, Longteng Guo, Jiang Jie and Jing Liu are with the Institute of Automation, Chinese Academy of Sciences, Beijing
100190, China, and also with the School of Artificial Intelligence, University
of Chinese Academy of Sciences, Beijing 100049, China. (e-mail: yuetongtian2022@ia.ac.cn; gaoxuange2022@ia.ac.cn;
xueshuning2021@ia.ac.cn; longteng.guo@nlpr.ia.ac.cn; jie.jiang@nlpr.ia.ac.cn; jliu@nlpr.ia.ac.cn). Yepeng Tang is with Beijing Jiaotong University, Beijing 100044, China
 (e-mail: yepengtang@bjtu.edu.cn) }}


\markboth{Journal of \LaTeX\ Class Files,~Vol.~14, No.~8, August~2021}%
{Shell \MakeLowercase{\textit{et al.}}: A Sample Article Using IEEEtran.cls for IEEE Journals}


\maketitle
\begin{abstract}

Electroencephalogram (EEG) signals are pivotal in providing insights into spontaneous brain activity, highlighting their significant importance in neuroscience research. However, the exploration of versatile EEG models is constrained by diverse data formats, outdated pre-training paradigms, and limited transfer learning methods, only leading to specialist models on single dataset or subject. In this paper, we introduce BrainGPT, the first generalist EEG foundation model designed to address these challenges. First, we propose an electrode-wise modeling strategy that treats the data of each electrode as an independent sample, enabling the integration of diverse EEG datasets collected from nearly all commonly used electrodes, amassing 37.5M pre-training samples. Second, we develop the first autoregressive EEG foundation model, moving away from traditional masked autoencoder approaches to a next-token-prediction task that better captures the temporal dependencies of EEG data. We also explore scaling laws with model up to billion level parameters — the largest scale in EEG research to date. Third, we introduce a multi-task transfer learning paradigm based on a learnable electrode graph network that is shared across tasks, which for the first time confirms multi-task compatibility and synergy. As a generalist EEG foundation model, BrainGPT shows broad compatibility with various signal acquisition devices, subjects, and tasks. It supports up to 138 electrodes and any combination thereof as input. Furthermore, we simultaneously evaluate it on 5 distinct downstream tasks across 12 benchmarks. BrainGPT consistently outperforms existing specialist models across all downstream tasks, with its effectiveness further validated through extensive ablation studies.
This work sets a new direction for generalist EEG modeling, offering improved scalability and adaptability for a wide range of EEG applications. Both the training code and model checkpoints will be publicly available.
\end{abstract}


\section{Introduction}
\IEEEPARstart{E}{lectroencephalogram} (EEG) captures spontaneous brain activity via electrograms, providing a non-invasive means to monitor neural dynamics \cite{biasiucci2019electroencephalography}. Through the analyses of EEG, valuable insights are derived for various applications, including but not limited to emotion recognition \cite{ye2022hierarchical}, motor imagery classification \cite{an2023dual}, mental workload detection \cite{wang2024arfn} and sleep stage classification \cite{liang2023teacher}. This breadth of applications underscores the versatility and utility of EEG in neuroscientific research. 

\begin{figure}[t]
\includegraphics[width=0.475\textwidth]{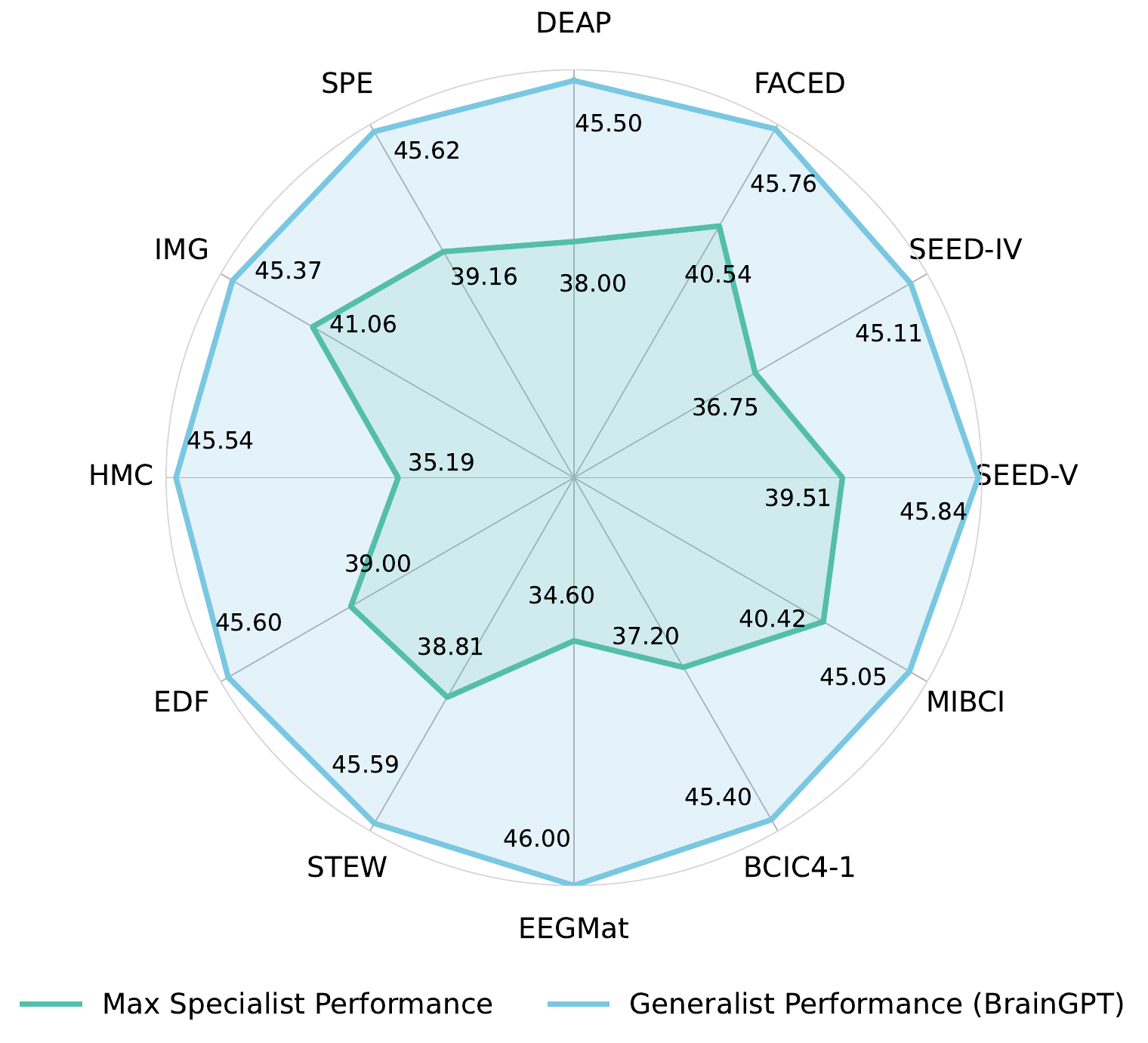}
\caption{BrainGPT, as a generalist model, significantly outperforms dataset-specific specialist models across 12 benchmarks spanning 5 tasks. It strongly demonstrate the versatility and transferability.}
\label{fig:radar_compare}
\end{figure}

Research on EEG downstream tasks has been thriving, yet most studies share a notable characteristic: \textbf{\textit{specialization}}. At the data level, for instance, a variety of proprietary data formats \cite{jia2020sst, bashivan2015learning} and handcrafted feature extraction techniques \cite{duan2013differential, yan2023topological} have been developed to enhance the discriminability of domain-specific data. Similarly, at the model level, numerous specialized modules and architectures are designed and optimized for specific tasks \cite{shao2023eeg}, datasets \cite{wang2024dmmr}, or even individual subjects \cite{gao2024multiscale}. While these approaches have advanced the field, they often lack generalizable and extensible capabilities. In contrast, an EEG model capable of simultaneously handling heterogeneous data and multiple tasks within a unified framework is highly anticipated. Such a model holds significant promise as a \textbf{foundation} for advancing EEG analysis by enhancing cross-domain adaptability, reducing the need for task-specific designs, and enabling more efficient learning.

Although some pioneering studies have made initial explorations into constructing EEG foundation models~\cite{jiang2024large, yang2024biot, yi2024learning}, three major challenges still remain:

\textbf{Data Format:} EEG data exhibit significant heterogeneity \cite{wang2024dmmr, saeed2021learning}, characterized by variations in data collection devices and experimental tasks. Furthermore, different datasets may employ a diverse number and combination of electrodes based on practical considerations. The inconsistency in data formats across different sources hinders their integrations into a unified model for training, posing a significant challenge in the development of an EEG foundation model. Therefore, an efficient and scalable strategy for unifying these diverse EEG data format is extremely demanding.

\textbf{Self-supervised Pre-training:} Current EEG foundation models~\cite{yang2024biot, jiang2024large, yi2024learning} predominantly rely on masked autoencoder (MAE) architectures~\cite{devlin2019bert, he2022masked}, which reconstruct masked portions of EEG signals using bidirectional attention mechanisms \cite{vaswani2017attention}. While effective in certain contexts, these approaches exhibit inherent limitations in modeling the temporal dependencies critical to EEG data. Specifically, MAE-based methods focus on reconstructing local signal segments, often neglecting the global temporal structure and long-range dependencies that are fundamental to understanding brain activity patterns \cite{song2023eeg, song2023global}. Considering that EEG signals often reflect a continuous and progressive flow of information, where past neural activity influences future states, the bidirectional nature of MAE architectures may not fully align this characteristic.

\textbf{Transfer Learning.} 
Current EEG foundation models are typically fine-tuned for specific datasets, resulting in specialists confined to narrow domains. However, this approach faces significant limitations in both efficiency and effectiveness. For efficiency, fine-tuning a separate model for each task is inherently costly, particularly given the increasing scale of models and the growing diversity of tasks. For effectiveness, such task-specific transfer fails to fully explore the potential for multi-task compatibility and synergistic interactions, thereby limiting the broader capabilities of foundation models.

Based on the above considerations, we propose BrainGPT, the first \textbf{generalist} EEG foundation model offering extensive versatility as shown in Fig.  \ref{fig:intro}. It seamlessly adapts and encodes signals collected by nearly all popular EEG acquisition devices, accommodating signals from up to 138 electrodes with various configurations and combinations. Moreover, it is capable of simultaneously supporting all prevalent downstream tasks within a single model, and it is highly scalable to new tasks. BrainGPT significantly diverges from previous methods, with its novelty encapsulated in three aspects: 

\begin{figure*}[t]
  \centering
  \includegraphics[width=0.9\textwidth]{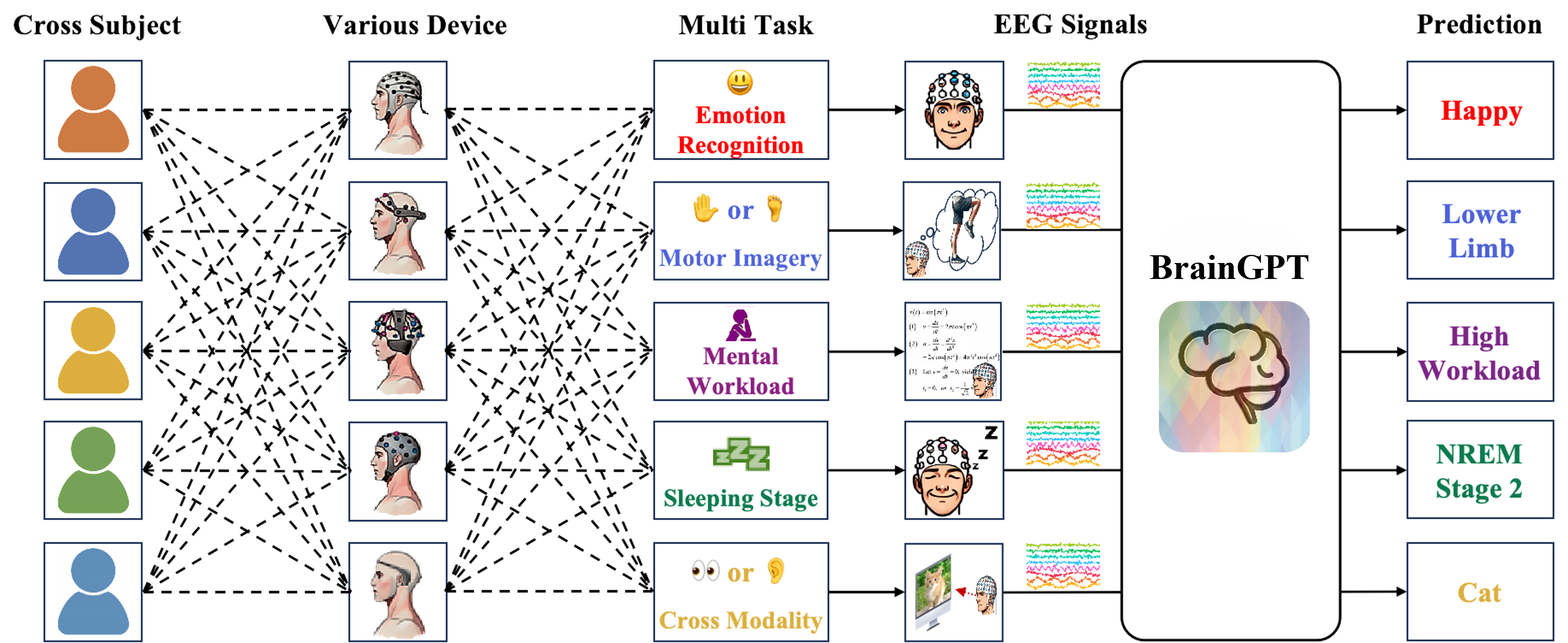}
  % \vspace{-12pt}
  \caption{The versatility of BrainGPT is reflected in the broad compatibility with subjects, signal acquisition devices, and tasks. EEG signals from various subject, using various device, and performing various task can be characterized and analyzed effectively within one model.}
  \label{fig:intro}
\end{figure*}

1) For data format, we propose an electrode-wise modeling strategy, which deconstructs EEG signals by treating each electrode as an independent data sample. It enables the model to learn the unique patterns and temporal dynamics of each electrode, with a learnable special token assigned to each electrode as prefix condition. Based on it, we collect a total of 37.5M pre-training samples with approximately 1B tokens.

2) For the pre-training paradigm, we propose the first autoregressive EEG model, which effectively captures the temporal dependencies inherent in each electrode through the intuitive yet challenging task of \textit{next token prediction}. We conduct pre-training across four scales (\textit{i.e.}, Base, Huge, Large, and Giant), with BrainGPT-Giant surpassing 1B parameters, making it the largest scale in the EEG field. Additionally, we firstly validate the scaling laws for both data and model size, confirming the scalability of this pre-training paradigm.

3) For transfer learning, we propose a learnable graph network with electrodes as nodes. It is concurrently shared across multiple tasks. Node activation patterns are adaptively determined by corresponding input task. Leveraging the robust temporal representations learned from autoregressive pre-training, the electrode graph serves as a spatial supplement by integrating information from multiple electrodes. The whole framework is designed with a progressive spatiotemporal decoupling. We collect data from 12 benchmarks for joint learning instead of traditional single-task fine-tuning. These tasks achieves promising multi-task compatibility and synergism, making BrainGPT the first generalist EEG foundation model.

Our contributions are summarized as follows:
\begin{itemize}
\item We introduce an electrode-wise modeling strategy, enabling flexible and scalable integration of heterogeneous EEG data. It allows BrainGPT to support up to 138 electrodes and arbitrary combinations, the largest coverage achieved so far, significantly enhancing generalizability.

\item We propose the first autoregressive pre-trained EEG model, which effectively captures the temporal dynamics inherent in EEG signals. We further validate the scaling laws of this pre-training paradigm with respect to data and model size, providing strong evidence for its scalability.

\item Building upon the task-shared graph network, significant mutual enhancement across tasks are demonstrated, making BrainGPT the first generalist model to exhibit confirmed multi-task compatibility and synergism.

\item BrainGPT demonstrates superior performance across 12 datasets encompassing 5 tasks, surpassing both pretrain-then-finetune and training-from-scratch predominant specialist baselines as shown in Fig. \ref{fig:radar_compare}. The effectiveness of our proposed method is further validated by extensive qualitative analyses.
\end{itemize}

\section{Related Work}

Despite recent exploratory studies on EEG foundation models gaining recognition, several challenges and limitations remain.
Existing studies \cite{yang2024biot, jiang2024large, yi2024learning, li2024temporal, wang2024cbramod, wang2025eegpt} exhibit notable commonalities and limitations in data formatting strategies, self-supervised pre-training paradigm, and transfer learning, which may constrain their applicability and scalability.

Existing studies typically segment EEG signals into fixed-length tokens, which are either flattened into "sentence"-like sequences or retain their spatiotemporal structure \cite{yang2024biot, jiang2024large, wang2024cbramod, wang2025eegpt}. To encode spatiotemporal information, fixed positional embeddings based on time or electrode location are commonly applied. While this formatting unifies diverse EEG inputs, it poses challenges.
Notably, many methods only utilize subsets of electrodes during pretraining \cite{yang2024biot, wang2024cbramod}, limiting scalability and generalization across datasets. The exclusion of signals from certain electrodes may impair performance on downstream tasks involving unseen channels.
To address this, we propose an electrode-wise modeling strategy, treating each electrode’s signal as an independent data sample. This enables flexible electrode combinations and leverages full-scale EEG data (up to 138 electrodes) during pretraining. 
% HERE

In terms of self-supervised pre-training paradigm, most existing methods predominantly adopt MAE techniques. For instance, BIOT \cite{yang2024biot} and MMM \cite{yi2024learning} utilize channel and temporal embeddings to construct EEG tokens for masked segment prediction, whereas LaBraM \cite{jiang2024large} employs a neural tokenizer to discretize EEG signals and predict masked tokens based on visible patches. 
However, these MAE-based method may destroy the natural flow of brain signals, making them less suited to the intrinsic temporal dependencies of EEG signals. Hence, in this paper, we refine the current pre-training paradigms from bidirectional completion to sequential prediction. To be specific, we propose the first autoregressive-based self-supervised learning framework for EEG foundation model pre-training, aiming to more effectively capture the temporal structure inherent in EEG signals.

In terms of transfer learning, existing models predominantly rely on computationally intensive task-specific adaptation strategies, where pre-trained models are fine-tuned for each downstream task \cite{yang2024biot, jiang2024large, li2024temporal} or even individualized per subject \cite{yi2024learning}. This approach leads to significant inefficiencies, particularly given the prohibitive costs of repeated full parameter optimization \cite{rafiei2022self}, and fails to leverage cross-task synergies inherent in EEG data. To address these limitations, we propose a unified task-shared transfer learning paradigm, enabling BrainGPT to function as a generalist foundation model capable of simultaneous multi-task compatibility and synergistic knowledge transfer without task-specific retraining.

In summary, while self-supervised pre-training has shown promise in EEG analysis, existing approaches exhibit constraints in data formatting, self-supervised learning paradigms, and transfer learning methodologies. Addressing these limitations is crucial for enhancing the generalization and performance of EEG foundation models, and the proposed BrainGPT provides a promising solution.

\section{Method}

\begin{figure*}[t]
  \centering  \includegraphics[width=\textwidth]{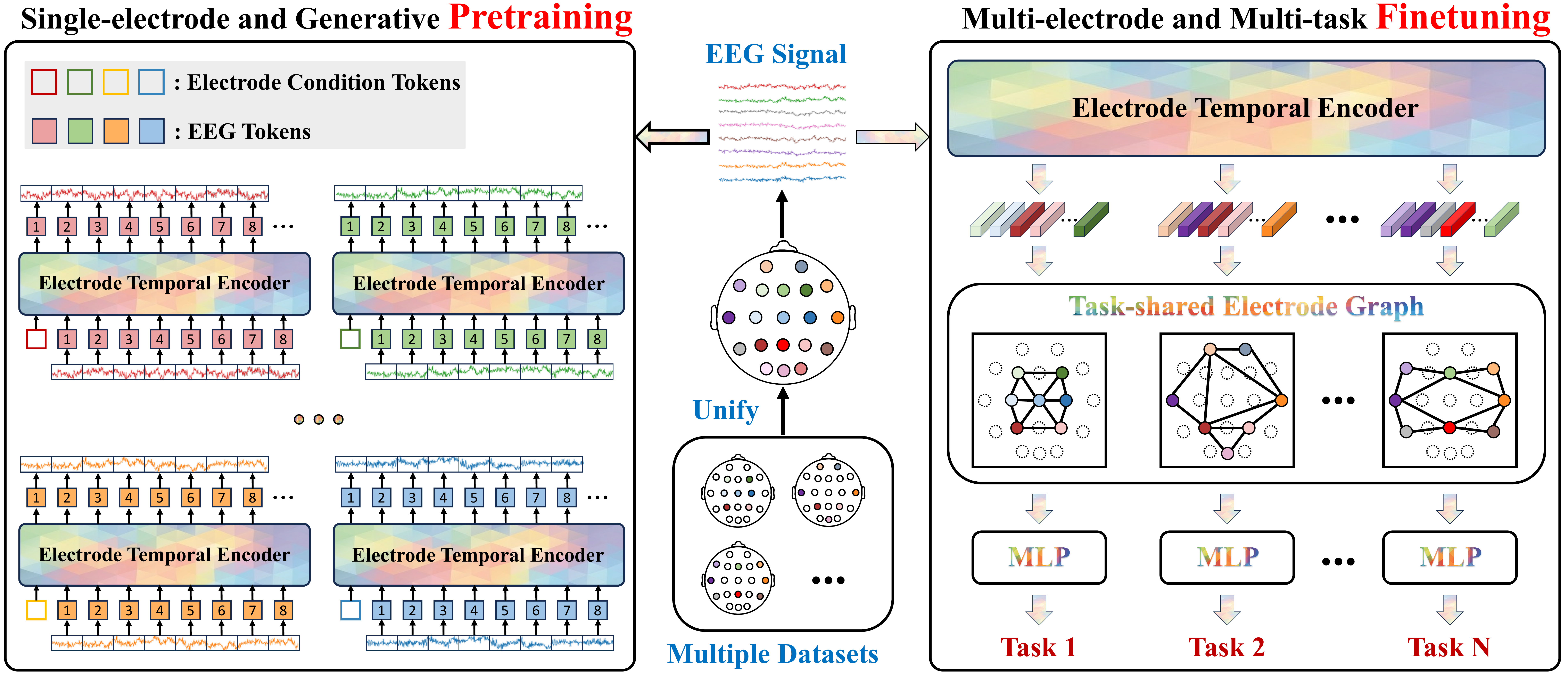}
  % \vspace{-12pt}
  \caption{Overview of the BrainGPT architecture. (left) Autoregressive reconstruction serves as the pre-training objective, with a learnable condition token added at the start to distinguish between electrodes. Each signal token predicts the next token through the Electrode Temporal Encoder (ETE) one-by-one. (right) Electrodes in each dataset are processed through the pre-trained ETE, extracting the final token as the electrode representations, which are then fed into a Task-shared Electrode Graph (TEG) network to integrate spatial information across multiple electrodes. ETE and TEG collectively constitute a progressive spatiotemporal decoupling.}
  \label{fig:methodology}
  % \vspace{-15pt}
\end{figure*}

In this section, we first introduce the electrode-wise modeling strategy adopted by BrainGPT, which decomposes the raw EEG signals on an electrode-wise basis to enhance flexibility. Subsequently, for each decomposed electrode signal, we propose a shared Electrode Temporal Encoder (ETE), which is pre-trained via autoregressive reconstruction to capture the temporal dynamics of individual electrodes. Finally, during the downstream task adaptation, we introduce the Task-shared Electrode Graph (TEG) network, designed to facilitate and enhance the collaborative learning of data across different tasks. The detailed framework is illustrated in Fig. \ref{fig:methodology}.

\subsection{Electrode-wise Modeling Strategy}

For a multi-electrode EEG signal, we first segment it into \( T \) one-second intervals. Each interval is represented by \( D \) uniformly sampled points from the original signal. We denote the segmented signal as \( x_i \in \mathbb{R}^{E_i \times T \times D} \), where \( E_i \) is the number of electrodes. Each EEG signal \( x_i \) is associated with an electrode set \( \mathcal{E}_i \) of size \( E_i \).  To facilitate electrode-wise modeling, we further decompose each \( x_i \) into individual electrode signals \( x_i^e \in \mathbb{R}^{T \times D} \). Each \( x_i^e \) represents the temporal sequence of a single electrode and serves as a basic training sample. The complete pre-training dataset \( X \) is then constructed as:  
\begin{equation}
X = \{ \mathbf{x}_i^e \mid e \in \mathcal{E}_i, \, i = 1, 2, \ldots, N \}
\end{equation}
The total number of electrodes covered in \( X \) is given by:
\begin{equation}
\label{eq:elec_num}
E = |\bigcup_{i=1}^N \mathcal{E}_i |.  
\end{equation}
To differentiate between electrodes, we introduce a trainable electrode vocabulary \( \mathcal{V} \in \mathbb{R}^{E \times D} \). All samples \( x_i^e \) from the same electrode share the corresponding embedding \( \mathcal{V}_e \). This embedding serves as a prefix condition and is concatenated along the sequence dimension with \( x_i^e \). For simplicity, we continue to refer to the concatenated sequence as \( x_i^e \).
\begin{equation}
   x_i^e = [\mathcal{V}_e||x_i^e] \in \mathbb{R}^{(T+1)\times D}
\end{equation}
 where $\|$ signifies the concatenation operation. Hence, signals from various domains and electrodes have been standardized into a highly scalable format. The chronological sequences $x_i^e$, which contains $\text{T+1}$ EEG \textit{tokens}, will serve as the fundamental unit for performing autoregressive reconstruction.

\subsection{Autoregressive Pre-training}

As shown in Fig. \ref{fig:methodology} (left), each \( x_i^e \) is fed into a shared Electrode Temporal Encoder (ETE), consisting of \( L \) identical layers. Each layer comprises two sub-layers: a multi-head \textit{causal} attention mechanism and a position-wise feed-forward network. Specifically, the input sequence \( x_i^e \) first passes through the \textit{causal} attention mechanism:
\begin{equation}
\text{Attention}(Q, K, V) = \text{softmax}\left(\frac{QK^T}{\sqrt{d}} + M \right)V
\end{equation}
\begin{equation}
Q =  W_Q x_i^e, \quad K =  W_K x_i^e, \quad V = W_V x_i^e 
\end{equation}
where $Q$, $K$, and $V$ are queries, keys, and values respectively, all derived from $x_i^e$ through corresponding learnable projection matrices $W_Q$, $W_K$ and $W_V$. The hidden size of ETE is denoted as $d$. $M$ is a causal mask designed to ensure that each token only attend to tokens that are sequentially prior to itself. The output of this sub-layer is then normalized and passed through a residual connection \cite{He_2016_CVPR}. Subsequently, it is fed into feed-forward network, which consists of two linear transformations with a Swish
\cite{dauphin2017language} activation:
\begin{equation}
    \text{FFN}(x) = W_{\text{down}} \cdot \left( \text{Swish}(\mathbf{W}_{\text{gate}} \cdot x) \odot (\mathbf{W}_{\text{up}} \cdot x \right)
\end{equation}
Finally, we employ a lightweight MLP to predict the next token in the sequence based on the output of ETE. The model is trained by minimizing the reconstruction loss, conditioned on all preceding tokens:  
\begin{equation}
\mathcal{L}(\theta) =  \frac{1}{T} \sum_{t=1}^T \rho( x^e_i[t] - \text{ETE}(x_i^e[\leq t]; \theta) )
\end{equation}
where \( \theta \) denotes the trainable parameters of ETE. The function \( \rho(\cdot) \) measures the discrepancy between the actual and reconstructed values, with Mean Squared Error (MSE) used as the default distance metric.

\subsection{Multi-task Transfer Learning}

Through autoregressive pre-training, ETE has effectively captured the temporal characteristics conditioned by electrodes. In this stage, unlike previous pre-trained models which are fine-tuned for individual tasks, we aim to explore a more versatile multi-task paradigm. Specifically, we propose a Task-shared Electrode Graph (TEG) network. This network adaptively activates interactions among various electrodes to simultaneously support multiple tasks.

Consider a multi-task dataset \(Y = \{y_1, y_2, \ldots, y_M\}\). Each sample \(y_j\) belongs to \(\mathbb{R}^{E_j \times T \times D}\), where \(E_j\) denote the number of electrodes. As illustrated in Fig. \ref{fig:methodology} (right), for each sample \(y_j\), a learnable special token \(c \in \mathbb{R}^{D}\) is broadcast across all \(E_j\) electrodes and appended to the end of the temporal sequence:
\begin{equation}
    y_j' = \left[ y_j || \ c \cdot \mathbf{1}_{E_j \times 1} \right] \in \mathbb{R}^{E_j \times (T+1) \times D}
\end{equation}
Leveraging the unidirectional attention mechanism inherent to autoregressive models, these special tokens facilitate the integration of local information to synthesize global representations for each electrode. $y_i'$ is then processed by the pre-trained ETE. Notably, the parameters of ETE are frozen during this stage, functioning solely as a feature extraction backbone. Subsequently, electrode representations are derived from the positions of special tokens in the output of ETE:
\begin{equation}
z_j = \left. \text{ETE}(y_j') \right|_{t=T+1} \in \mathbb{R}^{E_j \times D}
\end{equation}
Similarly, each sample in \(Y\) generates a corresponding \(z_j\) that captures comprehensive temporal information. These representations are then input into the proposed TEG network, which is specifically designed to model dependencies among electrodes, integrating spatial information effectively.

We initially construct a graph network in which each node represents an electrode utilized during the pre-training stage. The total number of nodes is the total number of electrodes $E$ in Eq. (\ref{eq:elec_num}). Benefiting from the comprehensive data coverage in the pre-training, these nodes include nearly all electrodes commonly employed, encompassing those found in $Y$. Each node is represented by a learnable vector of length $D$. They form a fully interconnected graph $\mathcal{G} \in \mathbb{R}^{E \times D}$. For each $z_j$, the electrodes it comprises are mapped to a subgraph $\mathcal{G}_j$ within $\mathcal{G}$.  Upon the introduction of \(z_j\) into the network, only the nodes contained within the subgraph \(\mathcal{G}_j\) are activated. This activation process involves updating the representations of these specific nodes by adding the corresponding elements from \(z_j\):
\begin{equation}
    \mathcal{G} = \mathcal{G}  + \mathbf{I}_{\mathcal{G}_j} \cdot \text{diag}(z_j) \cdot \mathbf{1}^T
\end{equation}
The function \(\text{diag}(z_j)\) converts \(z_j\) into a diagonal matrix, facilitating targeted activations that only influence corresponding nodes within \(\mathcal{G}_j\). The indicator matrix \(\mathbf{I}_{\mathcal{G}_j}\) ensures that updates are confined to these nodes, leaving others unaffected. The updated graph $\mathcal{G}$ facilitates the flow and interaction of spatial information between electrodes through a graph attention mechanism~\cite{velivckovic2017graph}:
\begin{equation}
\alpha_{mn} = \text{ReLU}(\mathbf{a}^T [\mathbf{W}h_m \| \mathbf{W}h_n])
\end{equation}
Specifically, for a given node \( m \), represented by \( h_m \in \mathbb{R}^{1\times D}\), the learnable projection weights \( \mathbf{W} \) and \( \mathbf{a} \) are used to compute the relevance $\alpha_{mn}$ to another node \( n \). The relevance scores are then normalized within all neighboring nodes $\mathcal{N}(m)$ of $m$ to obtain the attention coefficients:
\begin{equation}
\gamma_{mn} = \frac{\beta_{mn}  \cdot \exp\left(\alpha_{mn} \right)}{\sum_{k \in \mathcal{N}(m)} \beta_{mk} \cdot \exp(\alpha_{mk})}
\end{equation}
To ensure that interactions occur only within the activated subgraph \( \mathbf{I}_{\mathcal{G}_j} \), a masking coefficient \( \beta \) is introduced, where \( \beta_{mn} = 1 \) if both nodes \( m \) and \( n \) are within the subgraph, and \( \beta_{mn} = 0 \) otherwise. Based on the attention coefficients, the interactions between nodes are computed as follows:
\begin{equation}
h_m' = \text{ReLU}\left(\sum_{n \in \mathcal{N}(m)} \alpha_{mn} \mathbf{W}h_n\right)
\end{equation}
After interacting with nodes within the subgraph, the representation of node \( m \) is updated to \( h_m' \). Similarly, the above operation is stacked across \( K \) layers, with each layer employing a residual connection and pre-normalization. For various subgraphs \( \mathcal{G}_j \) (\textit{i.e.}, different datasets or tasks) within the same batch, unified training is efficiently achieved by constructing corresponding mask matrices $\beta$. This approach allows the model to operate as a multi-task generalist. In terms of output, the graph network pools the representations of nodes within \( \mathcal{G}_j \) and subsequently directs them to the relevant task-specific head for either classification or regression.

\section{EXPERIMENTS}

\subsection{Datasets}
We evaluate our BrainGPT using 12 datasets across 5 distinct tasks, as detailed in Table \ref{tab:benchmark}. For Emotion Recognition (ER), we utilize DEAP \cite{koelstra2011deap}, FACED \cite{chen2023large}, SEED-IV \cite{zheng2018emotionmeter}, and SEED-V \cite{liu2021comparing}. For Motor Imagery (MI) classification, we employ MIBCI \cite{cho2017eeg} and BCI Competition \MakeUppercase{\romannumeral4}-1 \cite{blankertz2007non}. For Mental Workload (MW) detection, we select EEGMat \cite{zyma2019electroencephalograms} and STEW \cite{lim2018stew}. For Sleeping Stage (SS) classification, we analyze data from EDF \cite{kemp2000analysis} and HMC \cite{alvarez2021haaglanden}. For Cross Modality (CM) tasks, we employed IMG, and SPE \cite{nguyen2017inferring}.

\begin{table}[t]
\centering
    \setlength{\tabcolsep}{4pt} % 缩小列间距
    \caption{Statistical analysis of 12 evaluation datasets.}
    \resizebox{\linewidth}{!}{ 
    \begin{tabular}{c|cccccc} 
    \toprule 
      {\bf Task} & {\bf Dataset} & {\bf Rate} & {\bf \# Subject} & {\bf \# Electrode} &  {\bf \# Sample} & {\bf \# Class}  \\
      \midrule
     \multirow{4}{*}{\color{ER} ER} & DEAP & 128Hz & 32 & 32 & 19.2k & 4 \\
     & FACED & 1000Hz & 123 & 30 & 27.6k & 9 \\
     & SEED-IV &  200Hz & 15 &  62 & 37.6k & 4 \\
     & SEED-V &   200Hz & 16 &  62 & 29.2k & 5 \\
     \midrule
     
     \multirow{2}{*}{\color{MI} MI} & MIBCI & 512Hz & 52 & 64 & 10.5k & 2 \\
     & BCIC4-1 & 100Hz & 7 & 38 & 1.4k & 2 \\
     \midrule
     
     \multirow{2}{*}{\color{MW} MW} & EEGMat & 500Hz & 34 & 19 & 1.0k & 2 \\
     & STEW & 128Hz & 45 & 14 & 3.3k & 3 \\
     \midrule
     
     \multirow{2}{*}{\color{SS} SS} & EDF & 100Hz & 78 & 2 & 19.5k & 5 \\
     & HMC & 256Hz & 151 & 4 & 22.6k & 5 \\
     \midrule
     
     \multirow{2}{*}{\color{CM} CM} & IMG & 1000Hz & 29 & 122 & 7.6k & 5 \\
     & SPE & 256Hz & 7 & 64 & 1.3k & 2 \\

    \bottomrule
    \end{tabular}}
    \label{tab:benchmark} 
\end{table}

For data preprocessing, all the EEG signals are resampled to 256 Hz. 
The signals are then filtered between 0.1 and 100 Hz and segmented into samples of four seconds.
Each sample is further segmented into 25 tokens with an overlap rate of 0.875, while each token has 256 sampling points.
Besides, we apply the z-score normalization.
No further preprocessing or artifact correction methods are applied. Across all datasets, we adopt a cross-subject paradigm. Specifically, we partition each dataset in the multi-task set into training, validation, and test splits using an 8:1:1 ratio, ensuring no overlap of subjects among these splits. To minimize variability, we calculate the average accuracy and standard deviation from results obtained using five distinct random seeds. 

\subsection{Baseline Models}
The baseline models selected for comparison are divided into two distinct categories. The first category encompasses traditional and widely utilized models in the EEG domain, such as EEGNet \cite{lawhern2018eegnet}, TSception \cite{ding2022tsception}, Conformer \cite{song2023eeg}, and LGGNet \cite{ding2023lggnet}. These models are trained from scratch on the respective datasets without  pre-training. The second category includes cutting-edge pre-trained models, \textit{i.e.}, LaBraM \cite{jiang2024large}, BIOT \cite{yang2024biot}. They are fine-tuned on the respective datasets using their publicly available pretrained weights. 
The detailed descriptions of the baseline models that we reproduce for comparison in this work are as follows:

\begin{itemize}
    \item \textbf{EEGNet} \cite{lawhern2018eegnet}: EEGNet is a compact convolutional neural network. It leverages depthwise and separable convolutions to facilitate feature extraction and classification.

    \item \textbf{TSception} \cite{ding2022tsception}: TSception is a multi-scale convolutional neural network, capable of learning discriminative representations across both time and channel dimensions. It incorporates a dynamic temporal layer to capture dynamic temporal and frequency representations, while an asymmetric spatial layer is employed to learn discriminative global and hemisphere representations.

    \item \textbf{Conformer} \cite{song2023eeg}: Conformer is a compact convolutional Transformer model, capable of encapsulating both local and global features. It incorporates a convolutional module to learn low-level local features while employing a self-attention mechanism to extract global correlations within the local temporal features.

    \item \textbf{LGGNet} \cite{ding2023lggnet}: LGGNet is a neurologically inspired graph neural network. It models the intricate relationships both within and between the brain's functional regions.

    \item \textbf{BIOT} \cite{yang2024biot}: BIOT is a self-supervised biosignal learning model that tokenizes biosignals of various formats into sentences. It can enable representation learning through a MAE-based pre-training strategy.

    \item \textbf{LaBraM} \cite{jiang2024large}: LaBraM is a self-supervised EEG model, which adopts the vector-quantized neural spectrum prediction to train a neural tokenizer that encodes EEG patches into compact neural codes. Through MAE strategy, it can learn generic representations.

\end{itemize}

Considering that no existing models have been evaluated on such a diverse range of downstream tasks, we have meticulously reproduced these models using their official code, hyperparameter configurations, and pretrained checkpoints. This reproduction aims to supplement the performance metrics for each task, facilitating a comprehensive comparison. It is important to note that all baseline results are derived from models fine-tuned for specific tasks, indicating that these are individual specialist models. In contrast, the results from BrainGPT originate from a single generalist model.

\subsection{Experimental Settings}

\subsubsection{Model Variants}

We have developed four architecture 
configurations of BrainGPT: BrainGPT-Base, BrainGPT-Large, BrainGPT-Huge, and BrainGPT-Giant. The parameter counts for these models are as follows: BrainGPT-Base is 1.46M, BrainGPT-Large is 11.29M, BrainGPT-Huge is 183.8M, and BrainGPT-Giant is 1.09B. In the case of the ETE and TEG network, they share the same hidden size and number of attention heads. These increments, which approximately scale by an order of magnitude at each level, are achieved by expanding the depth and width of the network.

\begin{table}[h]
\centering
    \caption{Configuration of BrainGPT models.}
    \resizebox{\linewidth}{!}{ 
    \begin{tabular}{@{}lccccc@{}}
    \toprule
     \bf Configuration &\textbf{Base}  & \textbf{Large} & \textbf{Huge}  &\textbf{Giant} \\ 
    \midrule
    ETE Layers & 3 & 9 & 12 & 20  \\
    TEG  Layers & 3 & 3 & 4  & 4  \\
    Head Size & 32 & 32 & 64 & 64  \\
    Hidden Size & 128 & 256 & 896  & 1,792 \\
    Attention Heads & 4 & 8 & 14  & 28  \\
    Intermediate Size & 512 & 1,024 & 3,584 & 7,168 \\
    \midrule
    Total Parameters & 1.46M & 11.29M & 183.8M & 1.09B  \\
    \bottomrule
    \end{tabular}
    }
    \label{tab:ablate_pretrain} 
\end{table}

\subsubsection{Training Details}

We adopt AdamW \cite{adamw} as the optimizer and conduct all training on 8 NVIDIA A800-SXM4-80G GPUs. To enhance training efficiency, we utilize DeepSpeed Zero Optimization Stage 2. During pre-training, all model scales are trained for 3 epochs using a consistent dataset of 37.5M samples, which collectively includes approximately 1B tokens. The batch size and learning rate are set to 4096 and 1e-4, respectively. For multi-task fine-tuning, all model scales are trained for 10 epochs using a consistent dataset of 181K samples. The batch size and learning rate are maintained at 512 and 1e-4, respectively. In this stage, the pre-trained parameters of ETE are frozen, and only the newly introduced TEG network is actively trained. 
The detailed training protocols of BrainGPT are reported in Table \ref{tab:hparam}, which contains the specific hyper-parameter configurations for the Stage I: Autoregressive pre-training and the Stage II: Multi-task fine-tuning.

\begin{table}[t]
\centering
\caption{Training hyperparameters for BrainGPT of two stages.}
\resizebox{1.0\linewidth}{!}{ 
\begin{tabular}{l|l|cccc}
\toprule
Stage & Hyperparameter & Base & Large & Huge & Giant \\ 
\midrule

\multirow{9}{*}{Stage I} & Lr & \multicolumn{4}{c}{1e-4} \\ 
                          & Epoch & \multicolumn{4}{c}{3.0} \\
                          & Precision & \multicolumn{4}{c}{BF16} \\
                          & Deepspeed & Zero2 & Zero2 & Zero2 & Zero3 \\
                          & LR Schedule & \multicolumn{4}{c}{cosine decay} \\
                          & Warmup Ratio & \multicolumn{4}{c}{0.03} \\
                          & Batch Size per GPU & \multicolumn{4}{c}{512} \\ 
                          & Gradient Checkpoint & \multicolumn{4}{c}{True} \\
\midrule
\multirow{9}{*}{Stage II} & Lr & \multicolumn{4}{c}{1e-4} \\ 
                          & Epoch & \multicolumn{4}{c}{10} \\
                          & Precision & \multicolumn{4}{c}{BF16} \\
                          & Deepspeed & Zero2 & Zero2 & Zero2 & Zero3 \\
                          & LR Schedule & \multicolumn{4}{c}{cosine decay} \\
                          & Warmup Ratio & \multicolumn{4}{c}{0.1} \\
                          & Batch Size per GPU & \multicolumn{4}{c}{64} \\ 
                          & Gradient Checkpoint & \multicolumn{4}{c}{True} \\
\bottomrule
\end{tabular}
}
\label{tab:hparam}
\end{table}

\subsection{Performance Evaluation}

\begin{table*}[t]
\centering
\caption{\textbf{Evaluation on EEG Benchmarks.} The column "One Model?" indicates whether the results for these benchmarks originate from the same model. The results in \textbf{bold} and \underline{underline} are the best and second-best results, respectively.}
\vspace{-2mm}
\resizebox{0.9\textwidth}{!}{%
\begin{tabular}{lccccccc}
\toprule
\multirow{2}{*}{Method} & \multirow{2}{*}{One Model?} & \multicolumn{4}{c}{\color{ER} Emotion Recognition}          & \multicolumn{2}{c}{ \color{MI} Motor Imagery} \\
\cmidrule(lr){3-6} \cmidrule(lr){7-8}
& & DEAP   & FACED  & SEED-IV & SEED-V  & MIBCI & BCIC4-1 \\
\midrule
\multicolumn{8}{l}{\textit{Specialist Models w/o pre-train}} \\
EEGNet & \ding{55} & 35.2\stdv{9.4} & 15.3\stdv{1.3} & 28.7\stdv{1.5} & 28.5\stdv{3.2} & 63.3\stdv{7.2} & 51.9\stdv{1.5} \\
TSception & \ding{55} & 34.3\stdv{8.1} & 14.0\stdv{1.8} & 32.2\stdv{3.6} & 29.9\stdv{7.0} & 61.4\stdv{6.5} & 52.2\stdv{1.6} \\
Conformer & \ding{55} & 38.0\stdv{8.7} & 14.1\stdv{3.6} & 29.6\stdv{2.3} & 26.5\stdv{1.0} & 52.6\stdv{3.0} & 51.6\stdv{1.8} \\
LGGNet & \ding{55} & 33.5\stdv{8.5} & 17.0\stdv{2.7} & 34.7\stdv{3.5} & 29.7\stdv{6.3} & 56.7\stdv{3.7} & 50.0\stdv{0.4} \\
\midrule
\multicolumn{8}{l}{\textit{Specialist Models w/ pre-train}} \\
BIOT & \ding{55} & 35.2\stdv{8.9} & 17.7\stdv{2.6} & 32.7\stdv{4.8} & 28.8\stdv{4.0} & 53.2\stdv{2.0} & -- \\
LaBraM  & \ding{55} & 34.3\stdv{9.9} & 15.5\stdv{1.6} & 29.5\stdv{2.1} & 26.4\stdv{0.7} & 50.5\stdv{1.1} & 50.3\stdv{0.4} \\
\midrule
\multicolumn{8}{l}{\textit{Generalist Models}} \\
EEGPT-Base & \ding{51} & 41.4\stdv{2.7} & 16.9\stdv{1.3} & 34.0\stdv{1.7} & 28.1\stdv{0.9} & 62.2\stdv{2.8} & 56.9\stdv{1.6} \\
EEGPT-Large & \ding{51}  & 42.5\stdv{3.8} & 17.8\stdv{1.7} & 36.3\stdv{2.1} & 30.1\stdv{3.7} & 63.4\stdv{4.4} & 57.3\stdv{1.0} \\
EEGPT-Huge & \ding{51}  & \underline{44.7\stdv{4.2}} & \textbf{20.7\stdv{2.3}} & \underline{38.7\stdv{1.9}} & \underline{32.3\stdv{2.7}} & \underline{65.7\stdv{2.6}} & \underline{59.1\stdv{1.3}} \\
EEGPT-Giant & \ding{51}  & \textbf{45.5\stdv{2.3}} & \underline{19.9\stdv{1.9}} & \textbf{41.3\stdv{1.5}} & \textbf{33.9\stdv{1.4}} & \textbf{67.2\stdv{3.3}} & \textbf{60.4\stdv{1.8}} \\
\addlinespace[0.5mm]
\midrule
\addlinespace[1mm]
\multirow{2}{*}{Method} & \multirow{2}{*}{One Model?} & \multicolumn{2}{c}{\color{MW} Mental Workload} & \multicolumn{2}{c}{\color{SS} Sleeping Stage} & \multicolumn{2}{c}{\color{CM} Cross Modality} \\
\cmidrule(lr){3-4} \cmidrule(lr){5-6} \cmidrule(lr){7-8}
&  & EEGMat  & STEW & EDF & HMC & IMG & SPE \\
\midrule
\multicolumn{7}{l}{\textit{Specialist Models w/o pre-train}} \\
EEGNet & \ding{55} & 60.0\stdv{8.7} & 52.3\stdv{17.6} & 84.0\stdv{4.4} & 54.5\stdv{8.7} & 38.1\stdv{5.1} & 52.2\stdv{1.4} \\
TSception & \ding{55} & 50.3\stdv{1.2} & 63.8\stdv{13.0} & 68.6\stdv{4.5} & 36.4\stdv{9.8} & 31.3\stdv{3.0} & 55.3\stdv{8.4} \\
Conformer & \ding{55} & 49.8\stdv{1.1} & 65.7\stdv{16.6} & 67.4\stdv{3.5} & 43.5\stdv{7.6} & 35.0\stdv{3.9} & 54.8\stdv{4.3} \\
LGGNet & \ding{55} & 50.2\stdv{1.1} & 46.7\stdv{12.5} & 68.6\stdv{4.5} & 17.0\stdv{9.5} & 34.5\stdv{3.5} & 52.4\stdv{5.8} \\
\midrule
\multicolumn{7}{l}{\textit{Specialist Models w/ pre-train}} \\
BIOT & \ding{55} & 50.2\stdv{1.1} & -- & -- & -- & -- & 53.4\stdv{4.9} \\
LaBraM & \ding{55} & 50.4\stdv{1.3} & 52.5\stdv{12.4} & 69.3\stdv{3.8} & 39.4\stdv{9.4} & 27.4\stdv{2.4} & 50.9\stdv{1.4} \\
\midrule
\multicolumn{7}{l}{\textit{Generalist Models}} \\
EEGPT-Base & \ding{51} & 66.0\stdv{8.6} & 63.2\stdv{10.6} & 85.2\stdv{3.4} & 65.5\stdv{4.0} & 38.1\stdv{1.9} & 58.2\stdv{2.6} \\
EEGPT-Large & \ding{51} & 69.0\stdv{3.5} & 65.4\stdv{10.1} & 89.0\stdv{2.2} & 66.8\stdv{3.5} & 39.2\stdv{2.5} & \underline{60.4\stdv{2.9}} \\
EEGPT-Huge & \ding{51} & \underline{70.7\stdv{6.2}} & \underline{68.5\stdv{12.8}} & \textbf{91.2\stdv{4.7}} & \underline{68.2\stdv{1.3}} & \underline{40.6\stdv{2.1}} & 60.3\stdv{3.5} \\
EEGPT-Giant & \ding{51} & \textbf{72.0\stdv{8.4}} & \textbf{70.7\stdv{11.9}} & \underline{90.6\stdv{1.9}} & \textbf{70.3\stdv{2.2}} & \textbf{41.5\stdv{1.7}} & \textbf{61.6\stdv{2.4}} \\
\bottomrule
\end{tabular}%
}
% \vspace{-4mm}
\label{tab:evaluation}
\end{table*}

Table \ref{tab:evaluation} presents a performance comparison across 12 datasets,  illustrating that BrainGPT, despite being a generalist model, consistently surpasses specialist models that have been fine-tuned for specific tasks. Specifically, BrainGPT-Giant achieves an average accuracy improvement of \textcolor{ER}{5.07\%} on the ER task, \textcolor{MI}{6.05\%} on the MI task, \textcolor{MW}{8.50\%} on the MW task, \textcolor{SS}{11.20\%} on the SS task, and \textcolor{CM}{5.10\%} on the CM task compared to the best performances by these specialist models. Moreover, as the model scales, there is a clear and consistent upward trend in performance improvement. 

Interestingly, our findings reveal that specialist models with pre-training appear to perform slightly worse than those trained from scratch. One intuitive hypothesis is that current mainstream EEG pre-training models are often based on large-scale seizure data, which exhibits domain discrepancy from typical EEG data used in general downstream tasks. This mismatch likely hampers the efficacy of transfer learning. Nonetheless, BrainGPT models demonstrate considerable versatility and effectiveness across a diverse array of tasks, thereby robustly validating its utility and performance.

\subsection{Ablation Study}

In this section, we conduct a detailed ablation analysis of the proposed training recipe. It is important to note that the findings are consistent across models of four different scales. Due to space limitations, we uniformly present the numerical results based on BrainGPT-Large.

\subsubsection{\textbf{Scaling law for model size preliminarily emerges}}

\label{sec:scaling_law}

\begin{figure*}[t]
  \centering  \includegraphics[width=1.0\textwidth]{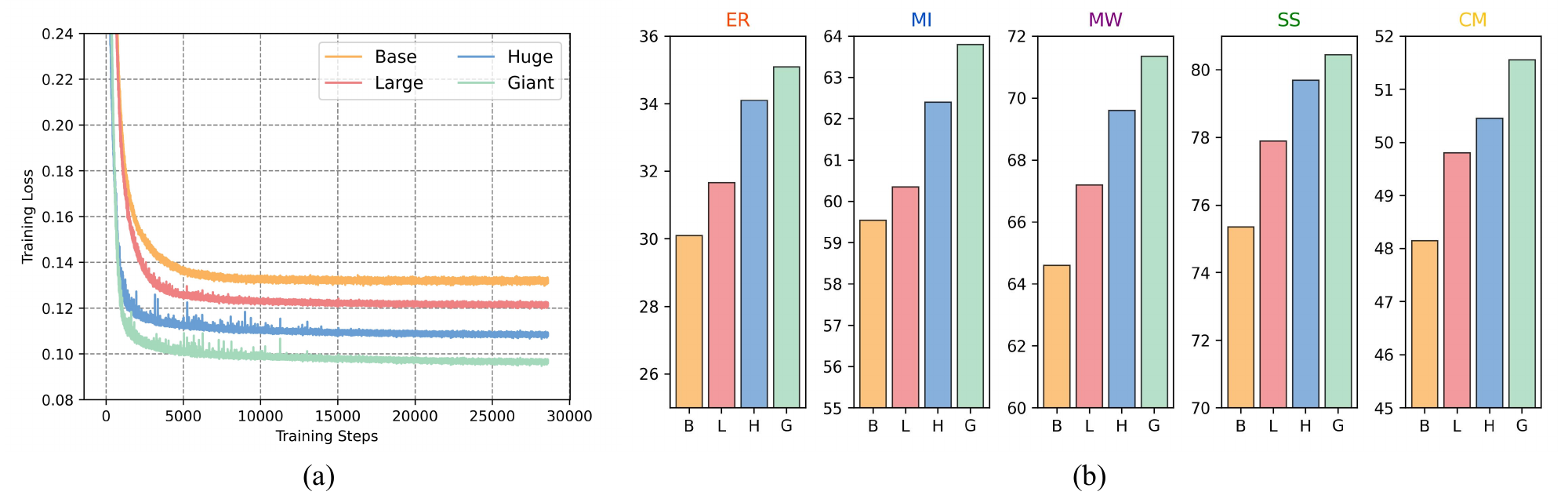}
  % \vspace{-12pt}
  \caption{Scaling laws for model size: (a) Pre-training loss curves of BrainGPT with varying parameter scales; (b) Performance of BrainGPT on 5 downstream tasks across different parameter scales.}
  \label{fig:scaling_law}
  % \vspace{-15pt}
\end{figure*}

Fig. \ref{fig:scaling_law} (a) compares the convergence curves of the autoregressive reconstruction loss across Base, Large, Huge, and Giant models. The results indicate that as the number of model parameters increases, the fit to the pre-training data improves, which is directly reflected in the final converged loss values. It largely indicates that models of a larger scale have absorbed more prior knowledge. We analyze the average performance of the four models across five tasks, as shown in Fig. \ref{fig:scaling_law} (b). For all tasks, a positive correlation between performance and model size is evident, suggesting that larger models can effectively transfer more pre-training knowledge to a wide range of downstream tasks. This represents the first effective exploration and validation of the scaling law for autoregressive models in the EEG domain. We believe that with further increases in training scale, autoregressive architectures may exhibit enhanced generalization and versatility.

\subsubsection{\textbf{Scaling law for training data preliminarily emerges}} 

\begin{figure*}[htbp]
  \centering  \includegraphics[width=1.0\textwidth]{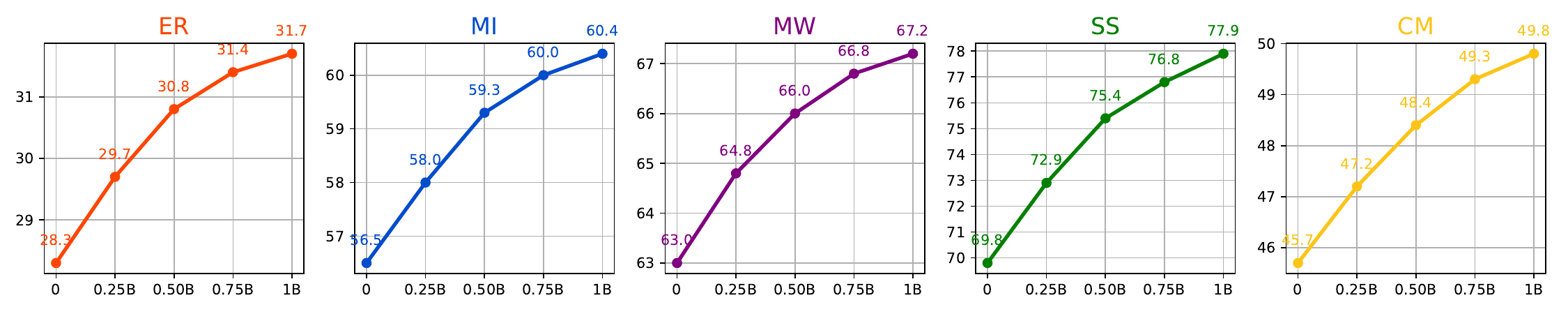}
  % \vspace{-12pt}
  \caption{Scaling laws for data volume: As the training data increases, performance improvements are observed consistently across 5 tasks.}
  \label{fig:data_scaling_law}
  % \vspace{-15pt}
\end{figure*}

In this section, we delve into another critical dimension: the scaling laws of training data. For our analysis, we randomly shuffle 1B tokens designated for pre-training and distribute them into five groups: 0B, 0.25B, 0.5B, 0.75B, and 1B tokens. Notably, the group with 0B tokens represents the absence of pre-training. We conduct pre-training across these varied data volumes. After freezing these pre-trained models, we perform multi-task fine-tuning, keeping training steps and settings consistent. The corresponding results are presented in Fig. \ref{fig:data_scaling_law}. As demonstrated in the figure, there are evident performance improvements across all five tasks as the volumes of pre-training data increase. These improvements are initially substantial but gradually taper off as data volumes expand. Similar patterns are also observed in the field of NLP \cite{kaplan2020scaling}. Given that the trend of performance improvement with increasing data has not yet diminished, we believe that by further expanding the dataset, BrainGPT could achieve even better performance.

\subsubsection{\textbf{Autoregression outperforms bidirectional masked pre-training}} 

Current EEG pre-training models employ a masked signal reconstruction framework using bidirectional attention \cite{jiang2024large, yang2024biot}. In this framework, random segments of the signal are masked, and the resulting input is processed by an encoder that reconstructs these segments based on contextual information (\textit{i.e.}, MAE). To enable a rigorous comparison between MAE and autoregressive (AR) modeling, we conduct an in-depth analysis using three distinct reconstruction loss functions: $\ell_1$, $\ell_2$, and cosine. For fairness, both pre-training paradigms utilize the same model architecture and parameter settings. We report related results in Table \ref{tab:ablate_ar}. For clarity, the standard deviations of the results presented have been omitted. The conclusions are twofold. First, for the three types of reconstruction loss, $\ell_2$ outperforms $\ell_1$, while $\cos$ shows the least efficacy. Second, irrespective of the distance metric employed, the AR architecture consistently outshines the MAE architecture, demonstrating a more than 2\% average accuracy advantage. These findings align with and support existing research in the NLP domain \cite{radford2019language, achiam2023gpt, brown2020language}. Specifically, the unidirectional modeling task poses significant challenges, enabling the model to learn more robust representations. Additionally, AR effectively adapts to the temporal characteristics of EEG signals, capturing their patterns more directly and naturally.

\begin{table}[t]
\centering
    \caption{Comparison of the models with the pre-training objective of MAE \textit{vs.} AR on 5 downstream tasks. Default setting is highlighted in \colorbox{lightblue}{blue}. }
    \resizebox{\linewidth}{!}{ 
    \begin{tabular}{c|c|cccccc}
        \toprule
        \textbf{Method} & \textbf{Loss} & \textcolor{ER}{\textbf{ER}}  & \textcolor{MI}{\textbf{MI}}  & \textcolor{MW}{\textbf{MW}} & \textcolor{SS}{\textbf{SS}} & \textcolor{CM}{\textbf{CM}} \\ 
        \midrule
        \multirow{3}{*}{MAE} & $\cos$ & 26.4 & 56.6 & 62.9 & 70.4 & 43.2 \\
        & $\ell_1$ & 27.8 & 60.8 & 61.6 & 73.2 & 45.3 \\
        & $\ell_2$  & 29.7 & 59.7 & 63.3 & 74.8 & 47.0 \\
        \midrule
        \multirow{3}{*}{AR} & $\cos$ & 28.6 & 59.1& 63.8 & 72.2 & 45.9 \\

        & $\ell_1$ & 30.0 & \textbf{61.2} & 65.0 & 74.5 & 48.6\\
        & $\ell_2$ & \cellcolor{lightblue}\textbf{31.7} & \cellcolor{lightblue}60.4 & \cellcolor{lightblue}\textbf{67.2} & \cellcolor{lightblue}\textbf{77.9} & \cellcolor{lightblue}\textbf{49.8}\\
        
        \bottomrule
        \end{tabular}
    }
    \label{tab:ablate_ar} 
\end{table}

\subsubsection{\textbf{Mutual enhancement is observed among various tasks}}

\begin{table}[t]
% \vspace{-11pt}
\centering
    \caption{Comparison of the models with the settings of  joint training \textit{vs.} separate training on 5 downstream tasks. Default setting is highlighted in \colorbox{lightblue}{blue}.}
    \resizebox{\linewidth}{!}{ 
    \begin{tabular}{c|ccccc}
        \toprule
        Settings & \textcolor{ER}{\textbf{ER}}  & \textcolor{MI}{\textbf{MI}}  & \textcolor{MW}{\textbf{MW}} & \textcolor{SS}{\textbf{SS}} & \textcolor{CM}{\textbf{CM}} \\ 
        \midrule
        separate  & 30.9 & 58.6 & 63.3 & 77.0 & 47.2\\
        \rowcolor{lightblue} 
        joint & 31.7\(\scriptstyle\textcolor{darkgreen}{\uparrow 0.8}\) & 60.4\(\scriptstyle\textcolor{darkgreen}{\uparrow 1.8}\) & 67.2\(\scriptstyle\textcolor{darkgreen}{\uparrow 3.9}\) & 77.9\(\scriptstyle\textcolor{darkgreen}{\uparrow 0.9}\) & 48.8\(\scriptstyle\textcolor{darkgreen}{\uparrow 1.6}\) \\
        \bottomrule
        \end{tabular}
    }
    \label{tab:sing-task} 
\end{table}

We compare two downstream task training settings—joint multi-task training (default) and separate training—as shown in Table \ref{tab:sing-task}. For clarity, the standard deviations of the results presented have been omitted. In the separate training setting, each model is trained independently for each task, utilizing the same number of iterations and architectures as in the joint training scenario. Our observations indicate that models utilizing joint training consistently outperform those with separate training across all five tasks. Actually, for the two different datasets, the corresponding subgraphs share overlapping nodes. These shared nodes (\textit{i.e.}, electrodes) provide a form of data augmentation that benefits both datasets. This augmentation is particularly important for tasks with limited samples. For instance, task \textcolor{MW}{MW} has a total size of only 4K, whereas task \textcolor{SS}{SS} reaches 42K. The accuracy benefits of joint training are more pronounced for task \textcolor{MW}{MW} compared to task \textcolor{SS}{SS} (\textit{i.e.}, \textcolor{MW}{3.9\%} for \textcolor{MW}{MW} \textit{vs.} \textcolor{SS}{0.9\%} for \textcolor{SS}{SS}). This enhancement suggests that despite originating from different tasks, signals from the same electrode exhibit shared patterns that can be effectively transferred. The introduction of the shared graph network effectively integrates and utilizes these shared patterns while also decoupling the differences between tasks. This phenomenon may provide an intriguing basis for future research on cross-task learning in EEG studies.

\subsubsection{\textbf{Generalized representational ability even on unseen data}}

\begin{figure*}[htbp]
  \centering  \includegraphics[width=0.9\textwidth]{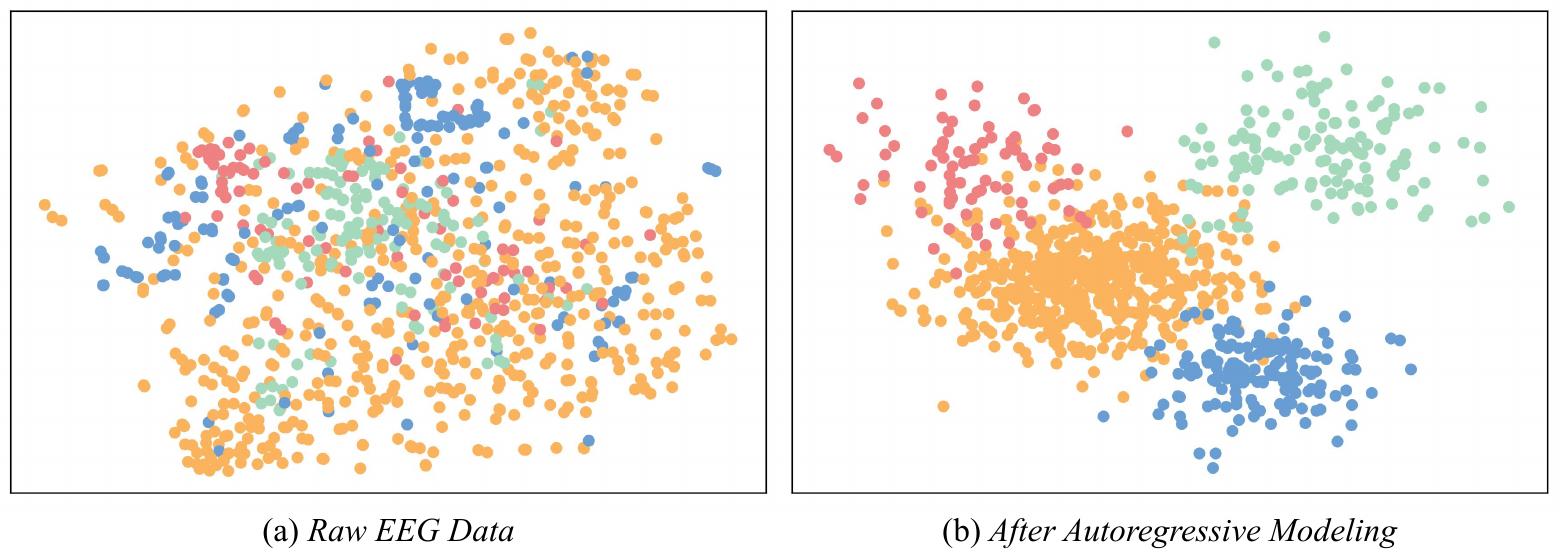}
  % \vspace{-12pt}
  \caption{t-SNE visualization comparison of the representation distributions before and after the autoregressive pre-trained Transformer. Different colors represent different categories.}
  \label{fig:tSNE_compare}
\end{figure*}

In this section, we explore an interesting conclusion regarding the transferability of ETE after autoregressive pre-training. Specifically, we utilize DREAMER \cite{katsigiannis2017dreamer}, a dataset which is not included during the pre-training stage. This dataset comprises four categories formed by the $2 \times 2$ combinations of high and low valence and arousal dimensions. It is fed into ETE to obtain representations. The entire process does not involve shared graph networks or require additional training. Consistent with the pre-training stage, for each electrode, we extract the last token as the global representation for that electrode. These last tokens are then averaged across electrode dimension, resulting in the final representation for each signal. We employ t-SNE \cite{tsne} to visualize the underlying structures and patterns within these representations, as illustrated in Fig. \ref{fig:tSNE_compare}. The findings indicate that autoregressive pre-training demonstrates strong transferability even on unseen data.

\section{CONCLUSION}
In conclusion, we have presented BrainGPT, the first generalist EEG foundation model designed to overcome the limitations of existing specialized EEG models. By introducing an electrode-wise modeling strategy, developing an autoregressive pre-training approach, and implementing a multi-task transfer learning paradigm with a learnable electrode graph network, BrainGPT unifies diverse EEG datasets and captures the sequential and temporal dependencies inherent in EEG signals. Our model demonstrates superior performance across 12 benchmarks, showcasing its versatility and scalability. We hope that BrainGPT will inspire further research and development in generalist EEG models.

\bibliographystyle{IEEEtran}
\bibliography{ref}

\end{document}